\documentclass{PoS}

\newcommand{\propover}{\raise0.5ex\hbox{\kern0.8em$\large\propto$\kern-1.35em\raise-1.5ex\hbox{\tiny{$p^2\to 0$}}}}
\newcommand{\toover}{\raise0.5ex\hbox{\kern0.8em$\large\to$\kern-1.35em\raise-1.5ex\hbox{\tiny{$p^2\to
0$}}}\kern0.5em}
\def\krto{ {\,\,\lower .8ex\hbox {$\longrightarrow \atop k \rightarrow 0$}\,\,}}

\def\bea{\begin{eqnarray} }
\def\beq{\begin{eqnarray} }

\def\eea{\end{eqnarray}}
\def\eeq{\end{eqnarray}}
\def\eq#1{eq.~(\ref{#1})}
\def\no{\nonumber \\ }

\title{A Ghost Story: Ghosts and Gluons in the IR regime of QCD}
\ShortTitle{A Ghost Story: Ghosts and Gluons in the IR regime of QCD}
\author{\speaker{Olivier P\`ene}\\
LPT Orsay (CNRS)~\footnote{Laboratoire de Physique
 Th\'eorique, Unit\'e Mixte de Recherche 8627 du Centre National de 
la Recherche Scientifique 
Universit\'e de Paris XI, B\^atiment 210, 91405 Orsay Cedex,
France}\\
E-mail: \email{olivier.pene@th.u-psud.fr}}
\author{Ph.~Boucaud\\
LPT Orsay (CNRS) \\
E-mail: \email{philippe.boucaud@th.u-psud.fr}}
\author{J.P.~Leroy\\
 LPT Orsay (CNRS) \\
E-mail: \email{jean-pierre.leroy@th.u-psud.fr}}
\author{A.~Le~Yaouanc\\
 LPT Orsay (CNRS)\\
E-mail: \email{alain.le-yaouanc@th.u-psud.fr}}
\author{J. Micheli\\
 LPT Orsay (CNRS) \\
E-mail: \email{jacques.micheli@th.u-psud.fr}}
\author{J.~Rodr\'iguez-Quintero\\
  Dpto. F\'isica Aplicada, Huelva~\footnote{Fac. Ciencias Experimentales,
Universidad de Huelva, 21071 Huelva, Spain} \\
E-mail: \email{jose.rodriguez@dfaie.uhu.es}}
\abstract{We discuss the different methods to obtain reliable informations
about the deep infra-red behaviour of the gluon and ghost Green functions in QCD.
We argue that a clever combination of analytical inputs and numerical ones is
necessary. We illustrate this statement about the distinction between two classes
of solutions of the ghost propagator Dyson-Schwinger equation (GPDSE). 
We conclude that the solution II (``decoupling") with a finite renormalised
ghost dressing function at zero momentum is strongly favored by lattice QCD,
We derive a
method to solve numerically the GPDSE using lattice inputs concerning the gluon
propagator. We derive an analytical small momentum expansion of the Ghost
dressing function. We prove from the large cut-off behaviour of the
ghost propagator renormalisation constant,  $\widetilde Z_3$, that the bare
ghost dressing function is infinite at the infinite cut-off limit.    
}
\FullConference{International Workshop on QCD Green's Functions, Confinement and Phenomenology\\
		 September 7-11 , 2009\\
ECT Trento, Italy}
\begin{document}
\begin{flushright}
{\small UHU-FP/08-010}\\
{\small LPT-Orsay/09-88}\\
\end{flushright}
\section{Introduction}
During the ten years ranging from 1965 to 1974 was invented the standard model 
of particle physics. This was a major and often overlooked scientific event. 
During the early sixties, the four fundamental interactions where known, but the
weak interaction was only described by Fermi's effective theory and the strong
interaction seemed to be even further away from any sound theoretical
description, precisely because, being strong, it seemed impossible to control 
by expanding around a small parameter, as had been the case for quantum
electrodynamics. 

Then came the miracle. A quantum field theory containing quarks and gauge
particles named gluons was proposed and its major property was isolated:
asymptotic freedom. It was a miracle because its formulation is extremely
compact, with only $n_f+1$ free parameters~\footnote{$n_f+2$ if we count the 
strong CP term.} naming $n_f$ the number of quark flavors, i.e. $n_f=6$, the
beautiful constraints of gauge symmetry, while the field of its applications is
huge. Up to now no strong argument has been presented which could allow to deny
QCD to be the theory of strong interactions. Of course there are drawbacks. The
first is that we are not able to extract very accurate predictions from QCD's
premises. 

But the most frustrating unsolved problem is the inexistence of a real proof of 
the confinement property, i.e. of the observation that only hadrons are observed
in nature and never isolated quarks or gluons. We are all convinced that
confinement is a property of QCD. Confinement is an experimental fact. 
Furthermore lattice-QCD (LQCD) calculations, which are based on QCD's principles,
 provide results in full agreement with confinement.  But this is not a proof.  
 
 Confinement is the major issue of this meeting and we all believe that it has
 to be looked for in the infrared behaviour of QCD. We will hear in this
 conference discussions around several approaches to confinement. Our approach
 will not be to follow or criticise some confinement scenarios, but rather to
 try to provide reliable answers to the question: How do Green functions behave
 in the deep infrared. In this talk we will, for the sake of simplicity,
  restrict ourselves to the quarkless pure Yang-Mills theory. We assume that
  the main features of QCD's infrared properties are present in Yang-Mills,
  at odds with Gribov's hypothesis that the light quark 
  supercritical binding was the origin of confinement~\cite{Gribov:1999ui}. 
 
 \subsection{Tools to handle the Green functions in the deep infrared}
 
 There exist analytical tools which are mainly Ward-Slavnov-Taylor  
 identities  (WSTI) and Dyson-Schwinger equations (DSE). There exists a numerical 
 tool which is LQCD. 
 
 WSTI and DSE are exact. They can be derived rigorously from the path integral 
 formulation of QCD. However, WSTI's are a necessary a posteriori check but do
 not constrain so much while DSE's are a very large
 set of coupled non linear integral equations. Trying to solve the latter is a
 formidable task and it is not clear how many solutions exist. There can
 even be an infinity of them. 
 
 LQCD is exact, it is really an approximation of QCD, however it is only
 numerical, leading to an intrinsic uncertainty, and, as we already mentioned,
 the accuracy is poor. 
 
 We believe that it is extremely fruitful to combine both the analytical and
 numerical approaches. Indeed we will see an example in which LQCD allows to 
 decide between two very different classes of solutions of DSE's. And next we
 will see an example in which analytic methods provide the functional form 
 of the ghost propagator in the deep infrared, thus allowing to extrapolate
 to zero momentum, where no direct LQCD calculation is possible.        
 
\subsection{Notations and definitions}

In latin languages the translation of ``ghost" starts with an ``F", while ``gluons"
starts with a ``G" in all languages we know. Therefore we use the following notations: 
the bare gluon propagator is written 
\bea
G_{\mu\nu}^{ab}(p^2,\Lambda^2) \equiv \frac {G(p^2,\Lambda^2)}{p^2}
 \;\delta_{ab} \,\left[\delta_{\mu\nu}-\frac{p_\mu\;p_\nu}{p^2} \right]
\eea
where $G(p^2,\Lambda^2)$ is the bare gluon dressing function, $\Lambda$ is the
ultraviolet cut-off, inverse lattice spacing $a^{-1}$ in the lattice case.
The bare ghost propagator is written
\bea
F^{ab}(p^2,\Lambda^2) \equiv \frac {F(p^2,\Lambda^2)}{p^2}
\; \delta_{ab}
\eea
where $F(p^2,\Lambda^2)$ is the bare ghost dressing function.

The corresponding renormalized quantities are labelled by the $R$ subscript:
\beq\label{defZ}
G_R(p^2,\mu^2) \equiv \lim_{\Lambda\to\infty} Z_3^{-1}(\mu^2,\Lambda) \;G(p^2,\Lambda^2)\qquad
F_R(p^2,\mu^2) \equiv \lim_{\Lambda\to\infty} \widetilde 
Z_3^{-1}(\mu^2,\Lambda)\; F(p^2,\Lambda^2)\qquad
\eeq
where $\mu$ is the renormalisation scale.

An important remark for the following is that~\cite{reisz1989,Luscher:1998pe}
\bea\label{cutoff}
F(0,\Lambda)
\ &=& \ 
\widetilde{Z}_3(\mu^2,\Lambda) 
\left( F_R(0,\mu^2) \ + \ {\cal O}\left(\frac 1 {\Lambda^2} \right) \right) \no
G(0,\Lambda)
\ &=& \ 
{Z}_3(\mu^2,\Lambda) 
\left( G_R(0,\mu^2) \ + \ {\cal O}\left(\frac 1 {\Lambda^2} \right) \right) \no
\eea

In the MOM renormalisation scheme, the renormalised quantities are set equal
 to their tree value when the momentum is equal to the 
 renormalisation scale:
 \beq
 G_R(\mu^2,\mu^2) =  F_R(\mu^2,\mu^2) \equiv 1 
 \eeq 
whence, using \eq{defZ}
\beq
Z_3(\mu^2, \Lambda^2) = G(\mu^2,\Lambda^2), \qquad \widetilde  Z_3(\mu^2, \Lambda^2)
 = F(\mu^2,\Lambda^2).
\eeq

The bare ghost-ghost-gluon vertex is parametrised by
\beq
\widetilde{\Gamma}_\nu^{abc}(-q,k;q-k) \ =  \
i g_0 f^{abc} \left( \ q_\nu H_1(q,k) + (q-k)_\nu H_2(q,k) \ \right) \ ,
\label{DefH12}
\eeq
where $k$ ($q$) is the incoming (outgoing) ghost momentum. Taylor's 
theorem~\cite{Taylor:1971ff}
implies that the ghost-ghost-gluon vertex becomes trivial when the incoming 
momentum vanishes
\beq\label{H1+H2}
H_1(q,0) +  H_2(q,0) = 1.
\eeq
This implies that if we take this kinematics to renormalise the ghost-ghost-gluon
vertex, the vertex renormalisation constant is $\widetilde z_1 = 1$.

Exploiting this property we can define very simply ``Taylor's coupling 
constant" by~\cite{vonSmekal:2009ae} 
\beq\label{alpha}
\alpha_T(p^2,\Lambda^2) = \frac{g_0^2(\Lambda^2)}{4\pi}\; F^2(p^2,\Lambda^2)\,
G(p^2,\Lambda^2).
\eeq
where $g_0^2(\Lambda^2)$ is the bare coupling constant.
Notice that, while $g_0, F $ and $G$ depend logarithmically on the 
cut-off $\Lambda$, $ \alpha_T$ only depends on it via inverse powers 
$O(1/\Lambda^2)$.

Finally we assume, as everybody does, some simple power law in the deep
infrared:
\beq\label{powers}
G(p^2,\Lambda^2)\; {\propover} \;(p^2)^{\alpha_G}\quad 
F(p^2,\Lambda^2)\; {\propover} \;(p^2)^{\alpha_F}\quad\Rightarrow\quad
\alpha_T(p^2,\Lambda^2)\;{\propover} \; (p^2)^{2\alpha_F+\alpha_G}
\eeq

\section{Two classes of solutions to the ghost propagator Dyson-Scwinger
equation}
Let us consider the ghost propagator Dyson-Schwinger
equation (GPDSE). It was claimed by many authors trying to solve the
DSE's that a general conclusion was that $2\alpha_F+\alpha_G=0$ or, in other
words that $ \alpha_T(p^2) \to$ ct $>0$ when $p^2 \to 0$. On the other hand many
indications from lattice QCD show a strong vanishing of $ \alpha_T(p^2)$, 
see a recent result at very small momenta in fig. 5 
of~\cite{Bogolubsky:2009dc}. 

Looking into details of the GPDSE we found that there were indeed two classes
of solutions~\cite{Boucaud:2008ky}

\begin{itemize}
\item Solution I: $2\alpha_F+\alpha_G=0,\, \alpha_T(p^2) \toover \mathrm{ct} >0$
and $\alpha_F <0 ,\; F(p^2,\Lambda^2) \toover \infty $
\item Solution II: $\alpha_F = 0,\; F(p^2,\Lambda^2) \toover  \mathrm{ct} >0 $
 and, using the lattice evidence that $\alpha_G >0$, $2\alpha_F+\alpha_G >0,\;
\alpha_T(p^2) \to 0$ 
\end{itemize}
This is valid at fixed cut-off $\Lambda$. Similar conclusions hold
for the renormalised quantities. Solution I is often called the ``scaling solution"
while solution II is called for some reason the ``decoupling solution".
\subsection{Schematic proof of the existence of the two solutions}

The GPDSE writes in our notations as
\bea
\label{SD1}
\frac{1}{F(k^2,\Lambda)}  = 1 + g_0^2 N_c \int \frac{d^4 q}{(2\pi)^4} 
\left( \rule[0cm]{0cm}{0.8cm}
\frac{F(q^2,\Lambda)G((q-k,\Lambda)^2)}{q^2 (q-k)^4} 
\left[ \rule[0cm]{0cm}{0.6cm}
\frac{(k\cdot q)^2}{k^2} - q^2  
        \right]
\ H_1(q,k,\Lambda)
           \right) \ ,
\eea
Using Taylor's theorem~\eq{H1+H2}, completed with indications from perturbative
QCD, we assume that in \eq{SD1} $H_1$ is regular: never vanishing nor infinite,
and not too far from 1. In practice we will take it as a constant close to 1.
This hypothesis is rather usual. The r.h.s of \eq{SD1} is divergent at fixed 
$\Lambda$ since the integrand $GF$ decreases at large $q^2$ as $\alpha^{35/44}$
which is not enough to make the integral convergent. Therefore  we prefer to 
regularize it by using a subtracted GPDSE:
\bea
\label{SD2}
\frac{1}{F(k^2)} - \frac{1}{F(k'^2)} &=& 1 + g_0^2 N_c \int \frac{d^4 q}{(2\pi)^4} 
\left( \frac{F(q^2)}{q^2 } \right) \Bigg( \rule[0cm]{0cm}{0.8cm} 
\frac{G((q-k)^2)}{(q-k)^4} 
\left[ \rule[0cm]{0cm}{0.6cm}
\frac{(k\cdot q)^2}{k^2} - q^2  
        \right]\no
&-&  \rule[0cm]{0cm}{0.8cm} 
\frac{G((q-k')^2)}{ (q-k')^4} 
\left[ \rule[0cm]{0cm}{0.6cm}
\frac{(k'\cdot q)^2}{k'^2} - q^2  
        \right] \Bigg)\ H_1
\eea
where we have assumed $H_1$ to be a constant and omitted to write $\Lambda$.
 The r.h.s. now is convergent.
Let us assume that we rescale all momenta by a common factor $\lambda \to
0$, we count the power behaviour of the l.h.s and the r.h.s. From \eq{powers}
the l.h.s behaves like $(\lambda^2)^{-\alpha_F}$ and the r.h.s as 
 $(\lambda^2)^{\alpha_G+\alpha_F}$. Matching both sides leads to solution I:
 $2\alpha_F+\alpha_G=0$.
 However, {\it there is a loophole in this argument when} $1/F(p^2)\to 1/F(0) >0$,
  i.e. when $\alpha_F=0$ since then the l.h.s. vanishes to leading order. To get
  a relation we need to go to the subleading behaviour of $1/F(p^2)$. We are
  then in the case of solution II: $\alpha_F=0$ and no constraint on
   $2\alpha_F+\alpha_G$. This is the proof. More details can be found 
   in~\cite{Boucaud:2008ky}.

\subsection{Numerical resolution of the GPDSE}
In order to understand better the relationship between these two classes of
solutions we have performed a numerical solution of the 
GPDSE~\cite{Boucaud:2008ji}. Since we consider only one DSE, we need additional
inputs. Our inputs are:
\begin{itemize}
\item The gluon propagator is taken from lattice QCD. It is extrapolated to the
large momenta using perturbative QCD formulae and to zero momentum assuming a 
finite, non zero limit, as strongly indicated by lattice QCD. 
\item The ghost-ghost gluon vertex is taken to be constant as justified above
from Taylor's theorem. 
\item The coupling constant multiplies the vertex function which we assume to be
constant. This product, a rescaled coupling constant, is taken as a free 
parameter.
\end{itemize}
We then fit this parameter to recover a solution in agreement with the ghost
propagator computed by lattice QCD. This exercize can be performed, mutatis
mutandis, with bare quantities or renormalized ones. In the latter case 
we define  the rescaled coupling constant by
\begin{equation}
\widetilde {g}^2 \equiv N_c g_{R}^2 \widetilde{z_1} H_{1R} 
\end{equation}
where $\widetilde{z_1}$ is the vertex renormalisation constant and our
renormalisation scale is chosen to be 1.5 GeV.

\vskip 0.5 cm
\begin{figure}[hbt]
\begin{center}
\includegraphics[height=7.8cm]{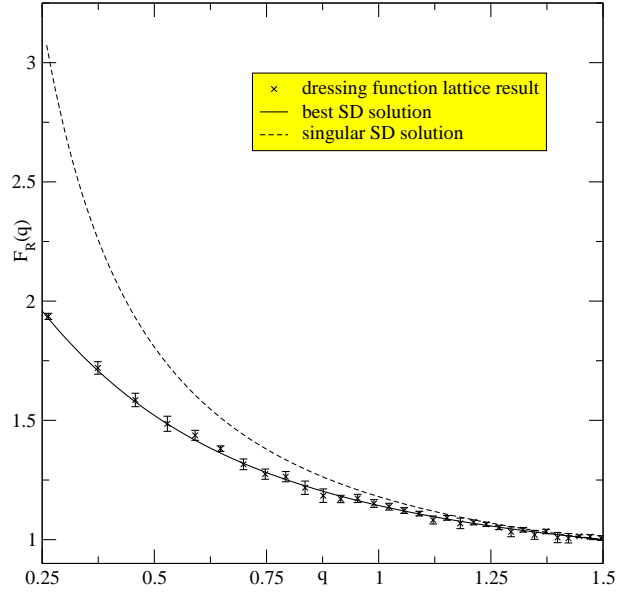}
\vskip 0.3 cm
\caption{\small Comparison between the lattice SU(3) data at $\beta=5.8$ and
with  a volume $32^4$ for the ghost dressing function and our continuum SD
prediction renormalised at $\mu=1.5$~GeV for $\widetilde g^2=29.$ (solid line) ;
the agreement is striking ; also shown is the singular solution which exists
only  at $\widetilde g^2=33.198....$ (broken line), and which is obviously
excluded.}
\label{fantomefig}
\end{center}
\end{figure}

\begin{figure}[hbt]
\begin{center}
\includegraphics[height=8cm,angle=-90]{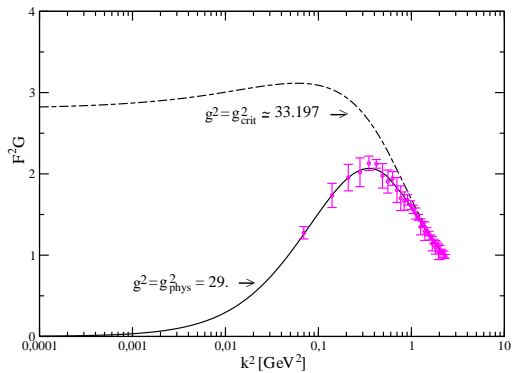}
\vskip -0.2 cm
\caption{\small Comparison between our lattice SU(3) data at  $\beta=5.8$ and
with  a volume $32^4$ for the product of gluon times  ghost square dressing
functions $G_R(k) F_R(k)^2$, renormalised at $\mu=1.5$~GeV, and the
corresponding curve for the continuum singular solution $\alpha_G+2 \alpha_F=0$,
which exists only   at $\widetilde g^2\simeq 33.198$, obviously excluded.  Also 
shown is our continuum regular solution for $\widetilde g^2=29$  (solid line)
for which the agreement is striking.}
\label{GF2fig}
\end{center}
\end{figure}

Our result is that {\it there is one critical value of the rescaled coupling
constant} $ \widetilde {g_c}^2 = 33.198 $ for which the renormalised ghost
dressing function diverges at zero momentum, solution I (``scaling"), while for 
all smaller $ \widetilde {g}^2$, $F(0)$ is finite, solution II (``decoupling").
Fitting to the values of $F(k^2)$ from lattice data gives $ \widetilde {g}^2 = 29
$. The plots are shown in fig.  \ref{fantomefig}. Not surprisingly, the plot
\ref{GF2fig} shows that the product $F^2G$, proportional to Taylor's coupling
constant, \eq{alpha}, goes to a constant for the critical $\widetilde {g_c}^2 $
and vanishes for any  smaller $\widetilde {g}^2$, fitting lattice data for $
\widetilde {g}^2 = 29 $.

\vskip 0.5 cm

\subsection{Expansion of the ghost propagator at small momentum}
\label{small}
From the GPDSE one can derive the low momentum expansion of the 
Ghost dressing function in the case of solution II~\cite{Boucaud:2008ji}.
If we assume that the gluon propagator goes to a finite constant, which implies
that $\alpha_T(k^2) \propto k^2$, it takes a simple form:
\bea\label{regular}
F_R(k^2,\mu^2)&\simeq& F_R(0,\mu^2)\left
 (1+\frac {3 \widetilde z_1 H_{1R}} {16 \pi} \alpha_T(k^2) \log(k^2)
\right)\no
F(k^2,\Lambda)&\simeq& F(0,\Lambda)\left 
(1+\frac {3 H_1} {16 \pi} \alpha_T(k^2) \log(k^2)
\right)\no
\eea

This formula, which can be refined~\cite{Next}, is very useful since it allows
an extrapolation of lattice data down to zero momentum. This is {\it an exemple 
in which an analytic method supports the numerical one}.

\section{What do we learn from lattice QCD}

This will be a very brief section as everything has been covered in Teresa
Mendes's talk. Let us just mention recent publications, which present 
results obtained with particularly large volumes and thus small momenta.
What follows concerns bare Green functions at some finite cut-off.
Cucchieri-Mendes have studied the $SU(2)$ case~\cite{Cucchieri:2008fc}: their 
fig.2 shows a bending of the ghost  
dressing function perfectly compatible with solution II (``decoupling"). 
In~\cite{Cucchieri:2009zt} they consider the $\beta=0$ situation and exhibit 
bounds on the gluon propagator (their fig. 4).  Bogolubsky et 
al.~\cite{Bogolubsky:2009dc} consider $SU(3)$: their fig. 2 shows that the gluon 
propagator goes to a non zero constant at zero momentum, fig. 4 shows also a 
 bending of the ghost dressing function 
and fig. 5 clearly shows a vanishing of $\alpha_T$ at zero momentum. The
  general conlusion is that the gluon propagator goes to a non zero constant, 
  the ghost dressing function may go to a finite non zero limit, and Taylor's
  coupling constant clearly vanishes at zero momentum. If the finiteness of the
  ghost dressing function is today only an indication, the vanishing of 
  the coupling constant is compelling, thus contradicting solution I
  (``scaling"). Now, since the analytic GPDSE method says that there exists only
  these two classes of solutions, we may conclude that nature has chosen
  solution II and thus that the ghost dressing function must indeed go to a finite
  non zero constant at finite cut-off.   
As we see {\it this is an exemple in which the LQCD numerical method allows to
discriminate between two classes of solutions of the GPDSE}.

One remark is in order here. We use the denomination ``coupling constant" in a 
very general sense: any well defined quantity which in the ultraviolet is
asymptotically equivalent to, say, $\alpha_{\overline {MS}}$, is eligible for
the denomination ``coupling constant". $\alpha_T$ defined in \eq{alpha} is
obviously one of those~\cite{vonSmekal:2009ae}. Is this coupling constant 
convenient for phenomenological descriptions using tree level diagrams in the
infrared ? presumably no. If one aims at this phenomenology, as do the proponents
 of the ``pinch technique"~\cite{Binosi:2009qm,Cornwall:2009ud}, one could
  easily redefine a new eligible coupling 
 constant by pulling a massive gluon propagator
out for the gluon leg amputation of the ghost-ghost-gluon Green function 
used to build the coupling~\cite{Aguilar:2009nf}. Thus $\alpha_{new}(k^2) = 
 \alpha_{T}(k^2) (k^2+M^2)/k^2$ where $M$ 
 could be the gluon mass\footnote{We thank A.C.~Aguilar, D.~Binosi, J.~Cornwall and
 J.~Papavassiliou for this comment.}. 

\section{ Can the bare ghost dressing function be finite non zero ?}

This question was raised by Kondo's remark~\cite{Kondo:2009ug,Kondo:2009gc} of 
a relation between the $k=0$ values of the  ghost dressing function $F(k)$,
Zwanziger's horizon function $h(k)$,  Kugo's function 
$u(k)$~\cite{Kugo:1979gm,Kugo:1995km}, and an additional
function $w(k)$. Applying to this relation Zwanziger's horizon gap equation and 
assuming that $w(0)=0$  he derives the surprising result that $u(0)=-2/3$ and
F(0)=3, independently of the cut-off. The questions we will raise are: is this
relation exact? does the prediction agree with lattice~? Is it compatible with
renormalisability of QCD ? Our point of view is detailed
in~\cite{Boucaud:2009sd}. 

\subsection{Kondo's relations}

In this subsection we only consider bare quantities.
One solution to the problem of Gribov's ambiguity,which was  
 proposed by Zwanziger~\cite{Zwanziger:1989mf},   
consists in using the Gribov-Zwanziger partition function, 
{\it which aims at restricting the Gribov copies~\cite{Gribov:1977wm}
 within the Gribov Horizon:}
\beq\label{GZ} Z_\gamma = \int \left[ D A \right]  \delta\left( \partial A \right) \ {\rm
det}(M) \ e^{-S_{\rm YM} + \ \gamma \int d^Dx h(x)} \ , \eeq
for the D-dimensional Euclidean Yang-Mills theory, 
where $S_{\rm YM}$ stands for the Yang-Mills action, $M$ is the 
Faddeev-Popov operator, 
\beq
M^{ab} = - \partial_\mu D^{ab}_{\mu} = - \partial_\mu \left( \partial_{\mu} 
\delta^{ab} + g f^{abc} A_\mu^c \right) 
\eeq
and $h(x)$ is Zwanziger's horizon function,
\beq
h(x)=  \int d^Dy \ g f^{abc} A_\mu^b(x) (M^{-1})^{ce}(x,y) g f^{afe}A_\mu^f(y) \ ;
\eeq 
that restricts the integration over the gauge group to the first Gribov region, provided 
that the Gribov parameter, $\gamma$, is a positive number. 

One  defines  then the function $u(k^2)$ which, at  vanishing 
momentum, gives the Kugo-Ojima parameter, and the function $w(k^2)$ 
via the following identities.
\bea\label{KOpar2}
\langle \ \left(D^{ab}_\mu c^b \right) \left(g f^{cde} A_\nu^d \overline{c}^e \right) \ \rangle_k
\ &=& -\ 
\delta^{T}_{\mu\nu}\delta^{ac} \ u(k^2) \ ;\no
\langle c^{a} \left( g f^{def} A_\nu^e \overline{c}^f \right) 
\rangle_k^{\rm 1PI}
& = & 
i \delta^{ad} \ k_\nu \left( u(k^2) + w(k^2) \right) \ .
\eea%

\begin{figure}
\begin{center}\vskip 8mm
\includegraphics[width=9.8cm]{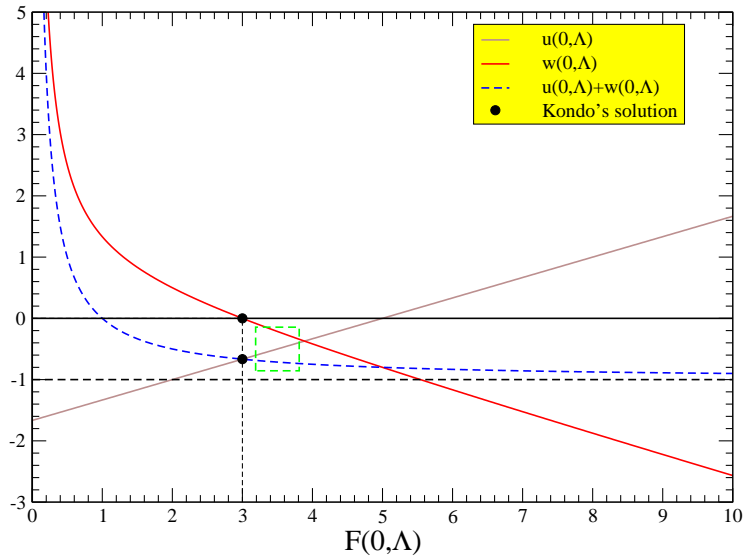}
\end{center}
\caption{ The solutions for $u(0,\Lambda)$ and $w(0,\Lambda)$  
plotted as a function of $F(0,\Lambda)$ under the assumption 
that the horizon gap equation is valid.}
\label{kondo-plot}
\end{figure}
From these definitions one obtains~\cite{Kondo:2009ug,Kondo:2009gc} 
and~\cite{Boucaud:2009sd} {\it without any 
hypothesis about $u$ and $w$}, 
\bea\label{solg}
u(0,\Lambda) &=& \frac{F(0,\Lambda)-1 }{D-1} \ -  \frac{D}{D-1} 
\left[\frac{\langle h(0) \rangle_{k=0}}{D(N^2-1)} \right]
 \\ 
w(0,\Lambda) &=& -1 - u(0,\Lambda) \ + \ \frac 1 {F(0,\Lambda)}\label{sols} 
 = - \frac{F(0,\Lambda)+(D-2)}{D-1} 
\ + \ \frac 1 {F(0,\Lambda)} \ + \frac{D}{D-1} 
\left[\frac{\langle h(0) \rangle_{k=0}}{D(N^2-1)} \right] \nonumber
\eea

\begin{figure}
\begin{center}\vskip 8mm
\includegraphics[width=9.5cm]{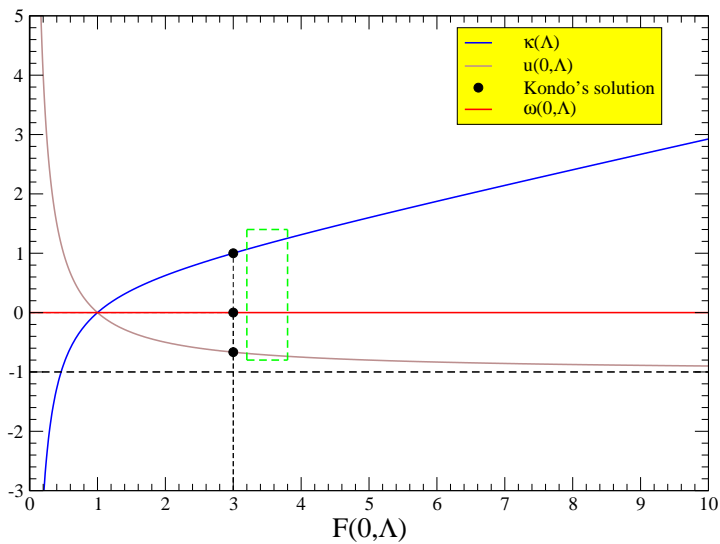}
\end{center}
\caption{ The same plot shown in fig.3 but 
$w(0,\Lambda)$ 
is required to be zero and the gap 
equation is relaxed by a multiplicative factor $\kappa(\Lambda)$,
 as explained in the text. $\kappa(\Lambda)$
is plotted on the solid blue line. 
Again, current lattice estimates lie inside the green dotted square.}
\label{w0-plot}
\end{figure}

\subsection{No finite $F(0,\Lambda)$ is possible at large cut-off $\Lambda$}

If we use Zwanziger's gap equation: 
\beq\label{horizon1}
\langle h(x) \rangle_\gamma = \left( N^2 -1 \right) \ D \ .
\eeq  
the functions $u(0)$ and $w(0)$ become, from \eq{solg},
 a function of the bare 
$F(0)$ plotted in Fig.~\ref{kondo-plot}.  
 The current  lattice solutions for the bare ghost dressing functions at
vanishing momentum lie inside the green  dotted square. {\it The  
apparent approximate agreement of lattice results with Kondo's solution 
is nevertheless misleading and
due to the moderate cut-off value on the lattices}.

Indeed let us assume a fixed renormalised $F_R(0,\mu^2)$. The plot
in fig.~\ref{kondo-plot} can then be understood, as a function of the 
$\widetilde Z_3(\mu^2,\Lambda)$ at fixed $\mu^2$ as soon as inverse powers of 
$\Lambda$ become negligible in front of logarithms, since, from 
\eq{defZ}, \eq{cutoff},  
\beq
\widetilde{Z}_3(\mu^2,\Lambda)= \frac{F(0,\Lambda)}{F_R(0,\mu^2)} + {\cal O}(\frac 1
{\Lambda^2}).
\eeq 
 The large cut-off dependance of $\widetilde{Z}_3$ is known to
be:
\beq\label{cutoffdep}
\frac{\widetilde{Z}_3(\mu^2,\Lambda)}{\widetilde{Z}_3(\mu^2,\Lambda_0)} \ = \  
\left( \frac{\log{(\Lambda/\Lambda_{\rm QCD})}}{\log{(\Lambda_0/\Lambda_{\rm QCD})}} \right)^{9/44} 
\ \left( 1+ \ {\cal O}\left( \alpha\right) \right)\ ,
\eeq
$\widetilde{Z}_3(\mu^2,\Lambda)\to \infty$ when $\Lambda\to\infty$.

Then the 
infinite cut-off limit is the limit at infinity on the horizontal axis of 
fig.~\ref{kondo-plot}.  The
particular solution proposed in ref.~\cite{Kondo:2009ug,Kondo:2009gc} (black
circles), obtained  by imposing $w(0,\Lambda)=0$, corresponds to the
intersection of $u+w$ and $u$. It cannot hold when $\widetilde{Z}_3\to \infty$. 
Notice that the hypothesis of a finite bare $F(0)$ with a vanishing $F_R(0)$ 
does not hold either since then $F(0,\Lambda) = \widetilde{Z}_3 F_R(0) +
{\cal O}(1/\Lambda^2)= {\cal O}(1/\Lambda^2)$
and consequently $F(0,\Lambda)$ vanishes when $\Lambda\to\infty$.  

Notice also from fig.~\ref{kondo-plot} that a finite $w(0,\Lambda)$ is not
possible at the large cut-off limit. 

We should now take into account that  gap equation (\ref{horizon1}) is a
consequence of Gribov-Zwanziger modification of the Yang-Mills action~\eq{GZ}.
This is not what is done in LQCD, although lattice gauge fixing also restricts
the Gribov copies within Gribov's horizon. Therefore we believe that 
condition \eq{horizon1} has no reason to be fulfilled in LQCD and maybe not at all in QCD. 
   Let us define  $\kappa(\Lambda)$ such that 
\beq\label{horizon1P} \langle h(0) \rangle_{k=0} =  \lim_{k \to 0} \frac 1 {V_D}
\int d^Dx \ \langle h(x) \rangle e^{i k \cdot x} \ = \ \kappa(\Lambda)\;  \left(N^2 - 1 \right) \
D \ . 
\eeq 
   
If the gap equation~\eq{horizon1} is thus relaxed  it becomes possible to 
keep $w(0)$ finite, a result derived in~\cite{Aguilar:2009pp,Grassi:2004yq} 
in the Landau background gauge. We show the solution when $w(0)=0$ on the 
fig.~\ref{w0-plot}. Nothing changes concerning the fact that the infinite 
cut-off limit is  at infinity on the horizontal axis. 
Our conclusion still remains valid:
{\it is not possible to have a finite $F(0,\Lambda)$ in the large $\Lambda$
limit}

\begin{figure}
\begin{center}
\vskip 8mm
\includegraphics[width=11.5cm]{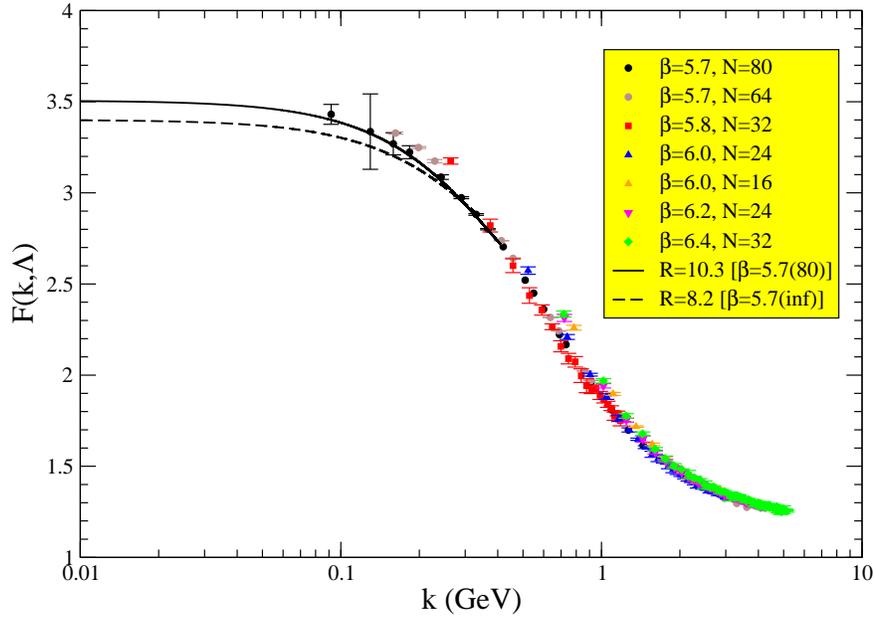}
\end{center}
\caption{Bare ghost dressing function estimated from different lattice 
data sets.  The solid line is for the best fit
with the small-momentum expansion with  $R(\beta=5.7(80^4))$ and
the dashed one stands for the best fit with
$R(\beta=5.7,\infty)$.}
\label{fig-ghosts}
\end{figure}

\section{Conclusion}
In fig.~\ref{fig-ghosts} we perform an extrapolation of the lattice bare ghost
dressing function using the small momentum expansion~\cite{Boucaud:2009sd}
 shortly explained in \eq{regular} of section  \ref{small}. The data for the two larger lattice volumes are taken from 
ref.~\cite{Bogolubsky:2009dc} and the others from 
refs.~\cite{Boucaud:2005gg,Boucaud:2008ji}. 
The fit with formula \eq{regular} is rather good in its range of validity (small
momentum). We also notice that the different lattice results seem to agree
rather well although they correspond to different $\beta$ values i.e. different 
lattice spacings. One may feel happy and claim that we have a good scaling
invariance. {\it But this is a wrong statement from a misleading observation}.
 
 Indeed, remember that the lattice spacing is the inverse of the
cut-off in lattice regularisation. From \eq{cutoffdep} we know that 
when the lattice spacing goes to zero, ($\Lambda \to \infty$), 
$F(0,\Lambda) \propto \beta^{9/44}$. On the whole range of lattice spacings
considered in fig.~\ref{fig-ghosts}, although the cut-off varies by more that a
factor of 3, $\beta^{9/44}$ varies only by 2.5\%. This is why this variation is
obscured by statistical errors~\footnote{Indeed there is a trend 
of the largest cut-off, $\beta=6.4$, to lie above the others, but this is hardly
visible.} in fig.~\ref{fig-ghosts}. This fake ``scaling invariance" hides the
truth: $F(k^2,\Lambda)$ rises very slowly to infinity when $\Lambda\to\infty$ 
i.e. $\beta\to\infty$. 

Bare values  depend dramatically, although slowly, on the cut-off and have
 no real meaning unless the
cut-off is specified. What makes really sense and has well defined limits at
infinite cut-off (vanishing lattice spacing) are renormalised
quantities~\cite{reisz1989,Luscher:1998pe}.   If we choose 1.5 GeV as the
 renormalisation scale, we get from lattice the gross estimate
\beq
F\,(1.5 \,\mathrm{GeV}) \equiv \widetilde{Z_3} \simeq 1.6 \quad\mathrm{whence}\quad
F_R\,(0,1.5 \,\mathrm{GeV}) \simeq 2.2.
\eeq    

Altogether, combining all which has been discussed here, our conclusion 
concerning the ghost dressing function is
\begin{itemize}
\item The renormalised ghost dressing function $F_R(0,\mu^2)$ has a finite limit
at vanishing momentum, $F_R(0,(1.5 \mathrm{GeV})^2)\simeq 2.2$. It is a positive
decreasing function at small momenta, probably  also decreasing for all 
momenta. 
\item The bare ghost dressing function $F(k^2,\Lambda)$ goes very slowly
to infinity  at infinite $\Lambda$ for all momenta.  
\end{itemize}


\begin{thebibliography}{99}


\bibitem{Gribov:1999ui}
  V.~N.~Gribov,
  Eur.\ Phys.\ J.\  C {\bf 10} (1999) 91
  [arXiv:hep-ph/9902279].
 
\bibitem{reisz1989}
T. Reisz, Nucl. Phys.B318 (1989) 417

\bibitem{Luscher:1998pe}
  M.~L\"uscher,
  arXiv:hep-lat/9802029.
 
\bibitem{Taylor:1971ff}
  J.~C.~Taylor,
  Nucl.\ Phys.\  B {\bf 33} (1971) 436.


      \bibitem{vonSmekal:2009ae}
        L.~von Smekal, K.~Maltman and A.~Sternbeck,
	     arXiv:0903.1696 [hep-ph].

\bibitem{Bogolubsky:2009dc}
  I.~L.~Bogolubsky, E.~M.~Ilgenfritz, M.~M\"uller-Preussker and A.~Sternbeck,
  Phys.\ Lett.\  B {\bf 676} (2009) 69
  [arXiv:0901.0736 [hep-lat]].

\bibitem{Boucaud:2008ky}
  Ph.~Boucaud, J.~P.~Leroy, A.~Le Yaouanc, J.~Micheli, O.~P\`ene and J.~Rodr\'\i guez-Quintero,
  JHEP {\bf 0806} (2008) 099
  [arXiv:0803.2161 [hep-ph]].

\bibitem{Boucaud:2008ji}
  Ph.~Boucaud, J.~P.~Leroy, A.~L.~Yaouanc, J.~Micheli, O.~P\`ene and J.~Rodr\'\i guez-Quintero,
  JHEP {\bf 0806} (2008) 012
  [arXiv:0801.2721 [hep-ph]].

 \bibitem{Next}
 Ph.~Boucaud et al. ``The low momentum ghost dressing function and the 
 gluon mass" in preparation.


\bibitem{Cucchieri:2008fc}
  A.~Cucchieri and T.~Mendes,
      Phys.\ Rev.\  D {\bf 78} (2008) 094503
        [arXiv:0804.2371 [hep-lat]].

\bibitem{Cucchieri:2009zt}
  A.~Cucchieri and T.~Mendes,
        arXiv:0904.4033 [hep-lat].

\bibitem{Binosi:2009qm}
 D.~Binosi and J.~Papavassiliou,
    Phys.\ Rept.\  {\bf 479} (2009) 1
      [arXiv:0909.2536 [hep-ph]].
      
\bibitem{Cornwall:2009ud}
  J.~M.~Cornwall,
    arXiv:0904.3758 [hep-ph].

\bibitem{Aguilar:2009nf}
 A.~C.~Aguilar {\it et al.},
 Phys.\ Rev.\  D {\bf 80} (2009) 085018
 [arXiv:0906.2633 [hep-ph]].
   
\bibitem{Kondo:2009ug}
  K.~I.~Kondo,
  Phys.\ Lett.\  B {\bf 678} (2009) 322
  [arXiv:0904.4897 [hep-th]].

\bibitem{Kondo:2009gc}
  K.~I.~Kondo,
  arXiv:0907.3249 [hep-th].

\bibitem{Kugo:1979gm}
  T.~Kugo and I.~Ojima,
  Prog.\ Theor.\ Phys.\ Suppl.\  {\bf 66} (1979) 1.
    
\bibitem{Kugo:1995km}
  T.~Kugo,
  arXiv:hep-th/9511033.

\bibitem{Boucaud:2009sd}
 Ph.~Boucaud, J.~P.~Leroy, A.~L.~Yaouanc, J.~Micheli, O.~P\`ene and J.~Rodr\'\i guez-Quintero,
     arXiv:0909.2615 [hep-ph].

\bibitem{Zwanziger:1989mf}
  D.~Zwanziger,
  Nucl.\ Phys.\  B {\bf 323} (1989) 513.

\bibitem{Gribov:1977wm}
  V.~N.~Gribov,
  Nucl.\ Phys.\  B {\bf 139} (1978) 1.
 
\bibitem{Aguilar:2009pp}
  A.~C.~Aguilar, D.~Binosi and J.~Papavassiliou,
  arXiv:0907.0153 [hep-ph].

\bibitem{Grassi:2004yq}
  P.~A.~Grassi, T.~Hurth and A.~Quadri,
  Phys.\ Rev.\  D {\bf 70} (2004) 105014
  [arXiv:hep-th/0405104].

\bibitem{Boucaud:2005gg}
  Ph.~Boucaud {\it et al.},
  Phys.\ Rev.\  D {\bf 72} (2005) 114503
  [arXiv:hep-lat/0506031];

 \end{thebibliography}
  \end{document}